\documentclass[11pt,letterpaper]{article}
\usepackage[T1]{fontenc}
\usepackage[utf8]{inputenc}
\usepackage{lmodern}
\usepackage[letterpaper,left=0.96in,right=0.96in,top=0.88in,bottom=0.95in]{geometry}
\usepackage{amsmath,amssymb}
\usepackage{graphicx}
\usepackage{array,booktabs,tabularx}
\usepackage{microtype}
\usepackage[font=small,labelfont=bf,labelsep=period]{caption}
\usepackage[section]{placeins}
\usepackage[round,authoryear]{natbib}
\usepackage{xurl}
\usepackage[hidelinks,unicode,pdfencoding=auto]{hyperref}
\setcitestyle{aysep={},yysep={;},citesep={;}}
\hypersetup{
  pdftitle={An Eight-Layer Framework for Synthetic ATC Speech Generation: Real-First ASR Adaptation and Operational Information Preservation at KDAB},
  pdfauthor={Juin Park and Chuyang Yang},
  pdfsubject={Airport-specific speech recognition and synthetic air traffic control speech}
}
\title{An Eight-Layer Framework for Synthetic ATC Speech Generation: Real-First ASR Adaptation and Operational Information Preservation at KDAB}
\author{Juin Park and Chuyang Yang\\[3pt]
  \small Embry-Riddle Aeronautical University, Daytona Beach, Florida, USA\\
  \small Corresponding author: Chuyang Yang (\href{mailto:yangc11@erau.edu}{yangc11@erau.edu})}
\date{}

\begin{document}
\maketitle
\begin{abstract}
Airport-specific automatic speech recognition (ASR) depends on verified local recordings, which are
costly to transcribe. An eight-layer framework generates synthetic air traffic control speech by
linking each operational scenario and reference transcript to its speech-synthesis settings and
radio-channel treatment. Using 963 verified transmissions from Daytona Beach International Airport
(KDAB), this study evaluates Whisper Large-v3 on 145 real test transmissions and measures
preservation of nine operational fields. In the original single-run comparison, adaptation on 673
real transmissions reduced word error rate (WER) from 66.84\% to 26.09\%. Adding 288 synthetic
examples (30\% of the training mixture) yielded 21.88\% WER, the lowest of the tested mixtures. In a
separate five-seed comparison on the same test set, the 30\% mixture lowered mean WER by 0.875
percentage points and outperformed real-only training in three of five pairs. In the original
comparison, full callsign preservation fell from 79 to 76 of 133 evaluable transmissions, and the
overall change in segment-level operational-field preservation was not statistically significant.
These findings support reporting both transcript accuracy and recovery of operationally critical
information when evaluating synthetic augmentation.
\end{abstract}

\section{Introduction}

Pilot-controller radio communication is an information-rich source of airport operational data
because it communicates clearances, requests, readbacks, handoffs, and position reports. Aircraft
movement alone cannot reveal all of this information. Recent accident investigations further
illustrate the importance of accurate communication and coordination in complex airport
environments. In its final report on the 2025 midair collision near Ronald Reagan Washington
National Airport (KDCA), the National Transportation Safety Board (NTSB) found that the use of
separate radio frequencies increased risk and that blocked transmissions prevented critical
instructions from being fully received \citep{NTSB2026}.

Reliable automatic speech recognition (ASR) could support communication archives, post-operation
review, and safety analysis while enhancing situational awareness among airport and aircraft
operators through communication-aware monitoring. However, operational usefulness depends on more
than general transcript readability. Operationally important fields such as callsigns, runway
identifiers, headings, and taxiways must be recovered precisely because a single altered digit may
associate an instruction with the wrong aircraft or change its meaning
\citep{Badrinath2022,Tanvir2026,Park2026}. This requirement is difficult to satisfy in
pilot--controller speech due to its fast-paced and highly condensed nature. Furthermore, the
specialized operational vocabulary is frequently transmitted over degraded radio channels prone to
noise and distortion. Consequently, these factors create a significant acoustic and linguistic
mismatch between real-world ATC audio and the broad speech data used to train general-purpose ASR
systems \citep{Helmke2021,Arra2024,Bagat2026}.

Airport-specific adaptation is constrained by the limited availability of manually verified local
speech. Recording availability does not translate directly into labeled training material: usable
transmissions must be identified within long audio streams, segmented, transcribed, and verified by
personnel familiar with aviation terminology. Poor radio quality, overlapping transmissions, and
ambiguous wording further reduce the amount of suitable ground truth, while access, redistribution,
and data-governance requirements may differ across recording sources. Privacy considerations may
also arise when recordings contain identifiable speaker information, depending on the recording
source and intended research use. Synthetic speech offers one way to extend limited real data by
introducing additional examples of aviation terminology, operational values, phrase structures, and
acoustic conditions. The remaining question is not simply whether synthetic speech can supplement
local recordings, but how much can be incorporated without weakening the authentic airport-specific
training signal. A related question is whether lower transcript error corresponds to better
preservation of operational information. WER and character error rate (CER) measure transcript
differences, but they do not distinguish between errors with different aviation consequences:
correcting an ordinary word can reduce WER without repairing an incorrect callsign, while a single
runway or heading error may alter the meaning of an otherwise accurate transcript.

This study examines whether operationally structured synthetic ATC speech can supplement limited
authentic tower communications from Daytona Beach International Airport (KDAB). Its primary
methodological contribution is an eight-layer framework that organizes the generation of
communication events and links operational content, phraseology, speaker delivery, and
radio-acoustic processing through traceable sample-level records. Using 963 manually curated
transmissions, it compares unadapted Whisper Large-v3, real-only Low-Rank Adaptation (LoRA), and
real--synthetic mixtures spanning 30\% to 70\% synthetic speech. The study contributes a structured
approach to low-resource synthetic supplementation, a controlled comparison of real-dominant through
synthetic-dominant training mixtures, and an evaluation combining transcript accuracy with
operational-information preservation. The research focuses on upstream ASR adaptation rather than
operational deployment or integrated ASR-ADS-B evaluation. More broadly, it addresses a practical
aviation and transportation gap: how to improve airport-specific ASR when verified local audio is
limited, and how to determine whether lower transcript error also preserves the information needed
to interpret aircraft identity and operational instructions.

\section{Related work}

Large aviation corpora and pretrained speech models have supported broader research on ATC
recognition and domain adaptation. The ATCO2 corpus provides speech collected across multiple
international airports, together with manually transcribed evaluation data and aviation-entity
annotations \citep{ZuluagaGomez2024}. \citet{Arra2024} showed that Whisper can be adapted using
public ATC corpora, while LoRA offers a computationally efficient means of modifying large
pretrained models \citep{Hu2022}. However, computationally efficient adaptation does not resolve the
availability of locally representative speech, and broad public corpora may not capture the
vocabulary, speaker characteristics, and radio conditions of a specific airport. In an earlier KDAB
evaluation, \citet{Park2026} compared four open-source ASR systems and identified Whisper Medium and
Large-v3 as the strongest candidates, although callsigns and intent-critical expressions remained
error-prone. These findings narrowed the current architecture screening to the two Whisper models
but did not resolve the underlying shortage of airport-specific training data \citep{Radford2023}.

Synthetic speech offers one way to expand that local training signal. \citet{Zheng2021} reported
improved recognition of underrepresented and out-of-vocabulary content when synthetic audio was
combined with authentic speech. \citet{Bagat2026} extended this approach to ATC by combining
text-to-speech, voice conversion, accent conversion, and acoustic simulation. Their synthetic-only
models improved over the unadapted baseline, and their best 50/50 real-synthetic condition achieved
lower WER than real-only fine-tuning. Their results demonstrate complementarity under a 50/50
design, but they do not determine whether an equal mixture is preferable in a low-resource,
single-airport setting. This unresolved composition question motivates the KDAB comparison of
real-dominant, balanced, and synthetic-dominant mixtures while holding the authentic training corpus
fixed. Complementing Bagat et al.'s emphasis on speech and acoustic variation, the present study
employs an eight-layer framework to structure operational communication content before speech
synthesis. \citet{Vu2026} examined the fidelity of synthetic ATC speech, including critical-token
preservation, whereas the present study evaluates the downstream effects of synthetic augmentation
on ASR performance using authentic KDAB recordings.

Evaluation must also account for the operational importance of individual transcription errors.
\citet{Helmke2021} showed that command-understanding performance and WER can diverge.
\citet{Arra2024} similarly reported differences between contextual and word-level evaluation, and
\citet{Arra2026} examined whether aviation entities such as callsigns, headings, altitudes, and
speeds remained represented despite transcription errors. These studies motivate a two-level
evaluation: WER and CER measure overall transcript accuracy, and operational-field preservation
measures whether aviation-relevant information survives. This two-level framework tests how changes
in synthetic share affect both overall transcription and the preservation of operationally
meaningful fields on the same fixed real KDAB test set.

Table~\ref{tab:t1} summarizes the methodological distinctions among selected studies of synthetic
speech augmentation and operationally informed evaluation, positioning the present work at the
intersection of structured ATC speech generation, airport-specific ASR adaptation, and
operational-information preservation.

\begin{table}[htbp]
\centering
\caption{Comparison of selected studies on synthetic speech augmentation and operationally informed ATC speech evaluation}
\label{tab:t1}
\footnotesize
\setlength{\tabcolsep}{4pt}
\renewcommand{\arraystretch}{1.13}
\begin{tabularx}{\linewidth}{@{}>{\hsize=0.65\hsize\linewidth=\hsize\raggedright\arraybackslash}X>{\hsize=1.16\hsize\linewidth=\hsize\raggedright\arraybackslash}X>{\hsize=1.03\hsize\linewidth=\hsize\raggedright\arraybackslash}X>{\hsize=1.16\hsize\linewidth=\hsize\raggedright\arraybackslash}X@{}}
\toprule
\textbf{Study} & \textbf{Primary research focus} & \textbf{Evaluation approach} & \textbf{Relationship to the present study} \\
\midrule
Zheng et al. (2021). & TTS-based augmentation for out-of-vocabulary recognition & Downstream ASR recognition performance & Demonstrates the value of synthetic training speech outside the ATC-specific setting \\
\addlinespace[2pt]
Helmke et al. (2021). & ATC speech recognition and command understanding & WER and command-level recognition measures & Establishes the importance of evaluating operational meaning beyond word-level accuracy \\
\addlinespace[2pt]
Bagat et al. (2026). & ATC synthetic-data generation using TTS, voice conversion, accent conversion, and acoustic simulation & WER after synthetic and real--synthetic ASR adaptation & Emphasizes speaker and acoustic diversification; provides a direct precedent for synthetic ATC augmentation \\
\addlinespace[2pt]
Vu \& Wei (2026). & Fidelity and operational plausibility of synthesized ATC speech & WER, critical-token F1, speaker similarity, and acoustic/\allowbreak{}prosodic measures & Evaluates generated speech quality rather than its downstream effect as ASR training data \\
\addlinespace[2pt]
Arra et al. (2026). & LLM-based entity-level assessment of aviation ASR transcripts & Aviation-entity representation evaluation & Provides a precedent for operationally informed ASR evaluation \\
\addlinespace[2pt]
Present study & Eight-layer operationally structured synthetic generation and airport-specific ASR adaptation & WER, CER, and nine-field operational preservation on authentic KDAB test recordings & Connects structured communication-event generation, real--synthetic mixture evaluation, and operational-information preservation \\
\bottomrule
\end{tabularx}
\par\smallskip
\begin{minipage}{\linewidth}
\footnotesize\raggedright
\textit{Note.} The comparison highlights selected methodological distinctions rather than offering an exhaustive review. Evaluation metrics and performance values are not directly comparable across datasets and experimental protocols.
\end{minipage}
\end{table}

\section{Methods}

\subsection{Real KDAB Corpus Construction}

This study used tower-radio communications from KDAB as the target domain for airport-specific
automatic speech recognition (ASR) adaptation. The real corpus was assembled from four broader
recording periods comprising ten 30-minute windows: 0200Z, 0230Z, 0300Z, 0330Z, 1230Z, 1300Z, 1700Z,
1730Z, 2200Z, and 2230Z. Sampling across multiple periods introduced variation in traffic activity,
speakers, communication content, and radio quality while retaining a common airport environment and
local operational vocabulary.

The recordings were manually reviewed and segmented at the individual-transmission level. Each
retained segment received a unique identifier and a human-verified reference transcript. The
associated manifest linked each segment to its source and audio metadata, verified transcript,
available speaker-role information, and the operational annotations used in subsequent evaluation.

\subsection{Preliminary ASR Benchmark and Base-Model Selection}

A preliminary benchmark was conducted to identify the base ASR model for the augmentation
experiments. Based on its overall transcription performance on the airport-specific ATC dataset,
Whisper Large-v3 was selected and used consistently in all subsequent experiments. \citet{Park2026}
report full descriptions of the candidate models, evaluation procedures, and comparative results.

\subsection{Structured Synthetic ATC Data Generation}

\subsubsection{Generation Framework}

\begin{figure}[htbp]
\centering
\includegraphics[width=1.0\linewidth]{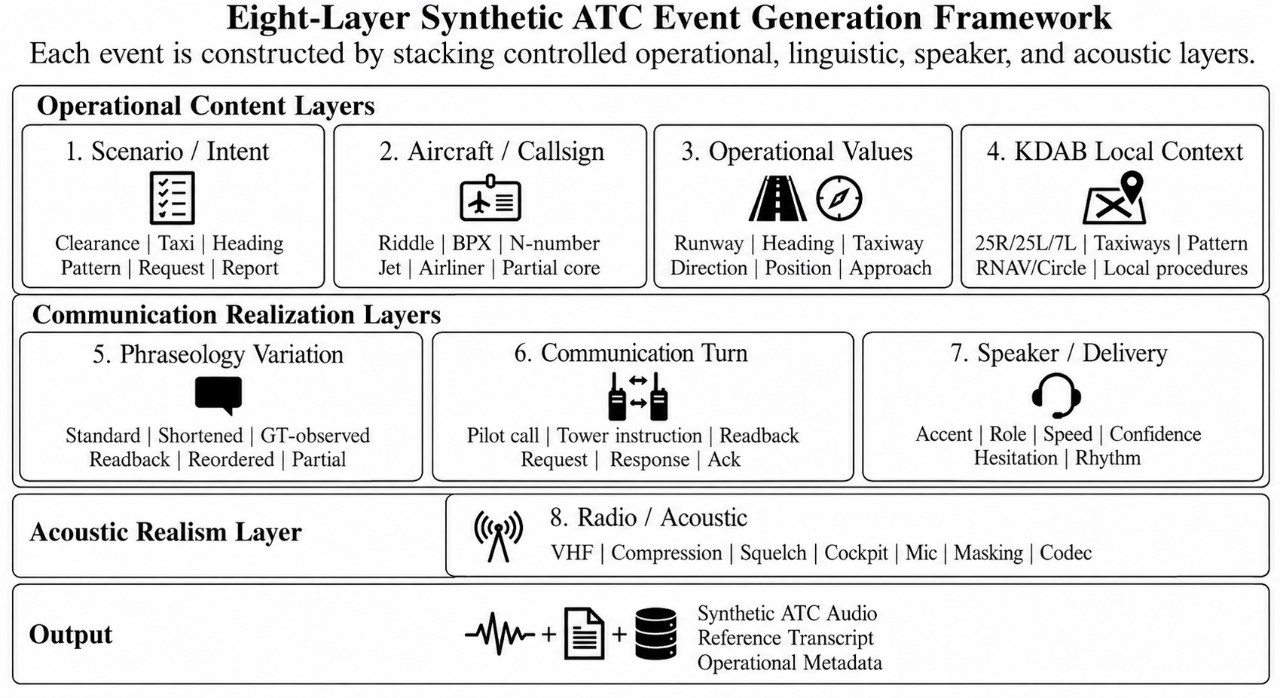}
\caption{Eight-layer synthetic ATC event-generation framework.}
\label{fig:f1}
\end{figure}

The synthetic corpus was designed as a structured communication-event dataset rather than as a
collection of independently generated aviation-sounding sentences. Each sample was constructed
through eight controlled layers: scenario or operational intent, aircraft identity and callsign,
operational values, KDAB-local context, phraseology variation, communication turn, speaker and
delivery condition, and radio or acoustic condition. The first six layers defined the intended
communication and reference transcript; the final two controlled speech realization and channel
characteristics. Each waveform remained linked to its reference transcript, operational fields,
synthesis settings, and acoustic-processing record through a common metadata identifier.
Figure~\ref{fig:f1} summarizes this eight-layer event-generation framework and the relationship
between transcript construction and audio realization.

\subsubsection{Operational Content and Transcript Construction}

Operational content was specified before the transcript was realized. Event-specific rules
determined the required fields: heading instructions required a heading value and, when applicable,
a turn direction; landing and takeoff clearances required a runway and clearance type; taxi events
required surface-movement information; and inbound or position reports required a location, pattern,
or approach context. Callsign identity was represented separately from sentence structure, allowing
controller calls, pilot requests, full readbacks, and shortened readbacks to use role-appropriate
placement and completeness.

KDAB-oriented value pools constrained runways, taxiways, approaches, traffic-pattern terms, and
locally recurring callsign families. Transcript generation followed a standard-first hierarchy in
which the communication event and speaker role selected a base template, followed by controlled
variation in wording, callsign placement, readback completeness, and operationally plausible
nonstandard phrasing. This process produced a traceable pool of 2,000 synthetic scenarios for
mixture-ratio experiments and component-level sensitivity analysis.

\subsubsection{Speech Synthesis and Delivery Control}

Clean speech was generated with Google Cloud Text-to-Speech Standard voices. The final
2,000-scenario pool used multiple en-US Standard voices for 1,600 samples (80\%) and en-IN Standard
voices for 400 samples (20\%). Voice selection was deterministic within predefined voice pools, and
the synthesis manifest retained the language code, voice identifier, speaker role, delivery profile,
and speaking rate assigned to each scenario. This distribution was a controlled voice-locale design
rather than an estimate of the demographic or accent composition of KDAB radio users, for which no
authoritative speaker-level data were available.

The en-US voice pool served as the primary U.S.-airport anchor. The en-IN condition was included
because Indian-influenced speech was encountered repeatedly during manual review of the KDAB
recordings, and dedicated en-IN Standard voices were available. Some scenarios initially assigned
Chinese- or Spanish-influenced metadata labels were synthesized with en-US Standard voices and,
therefore, were not treated as distinct accent realizations; their contribution was limited to the
broader voice, role, and delivery variations. Speech was generated as 24-kHz LINEAR16 audio using
role-aware speaking rates of 0.85x, 0.90x, 0.95x, 1.00x, 1.10x, 1.20x, and 1.30x, with 0.95x as the
default. Minimal Speech Synthesis Markup Language was used, including a 500-ms final break to reduce
end-of-utterance truncation.

Speaking-rate variation was implemented as a controlled TTS delivery parameter and did not
explicitly reproduce spontaneous speech reductions, such as syllable elision and
coarticulation-dependent shortening, that may occur in rapid ATC communication. Consequently, the
configured speaking rates should not be interpreted as validated representations of naturally
accelerated KDAB speech.

\subsubsection{Radio and Acoustic Processing}

After synthesis, the clean audio files were resampled to 22,050 Hz and processed using predefined
radio-style channel and interference treatments. These treatments were not calibrated to measured
KDAB background noise or receiver characteristics.

\begin{figure}[htbp]
\centering
\includegraphics[width=0.94\linewidth]{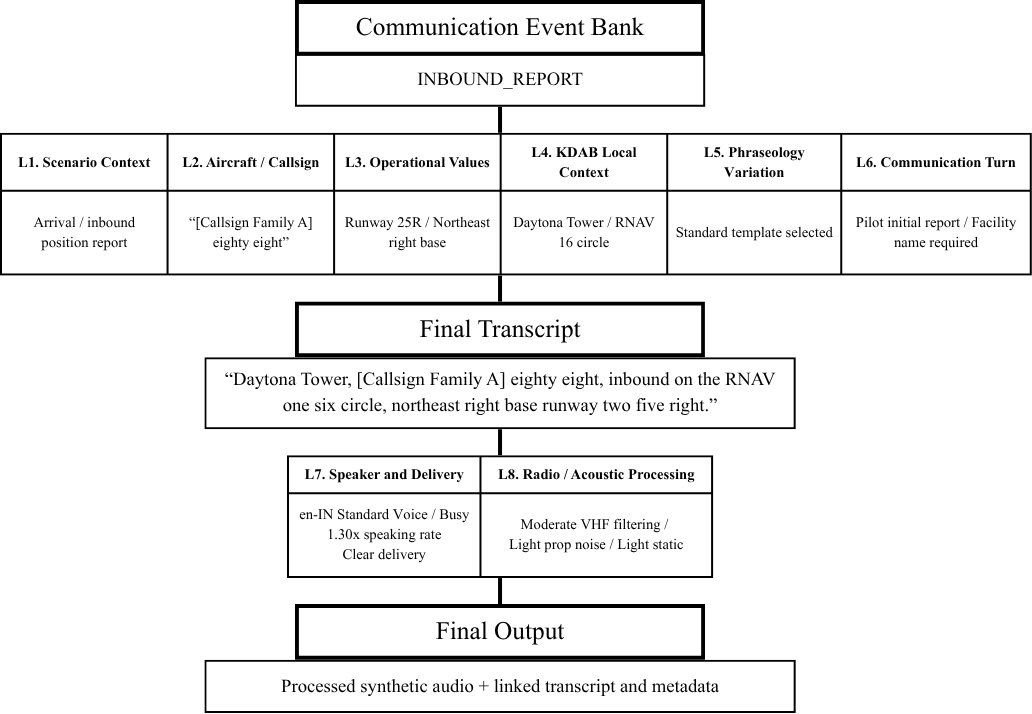}
\caption{Worked example showing the application of the eight-layer synthetic ATC event-generation framework to a representative KDAB inbound-report scenario.}
\label{fig:f2}
\end{figure}

Seven preset acoustic profiles were used, ranging from lightly filtered radio speech to noisier
cockpit, weak-signal, and clipped-microphone conditions. Most samples received light or moderate
processing, while the more severe conditions were used less frequently. Each processed file remained
linked to its transcript, voice settings, operational metadata, and assigned acoustic profile.
Figure~\ref{fig:f2} illustrates the application of the eight-layer framework through a
representative KDAB inbound-report scenario, showing how the selected event, callsign, operational
values, local context, phraseology, communication turn, speaker-delivery condition, and acoustic
profile were combined into a traceable synthetic sample.

\subsection{ASR Adaptation and Experimental Conditions}

Whisper Large-v3 was evaluated under five principal conditions: an unadapted baseline, real-only
LoRA adaptation, and real-plus-synthetic mixtures with synthetic shares of 30\%, 50\%, and 70\%.
Four additional models (G--J) were evaluated in component-level sensitivity analyses relative to the
30\% mixture. Table~\ref{tab:t2} summarizes the training compositions and distinguishes the
principal A--E comparisons from the exploratory G--J conditions. The real training contribution
remained fixed at 673 segments, and synthetic counts were calculated as a proportion of the complete
training mixture.

\begin{table}[htbp]
\centering
\caption{Integrated ASR Experimental Conditions and Main-Analysis Membership}
\label{tab:t2}
\footnotesize
\setlength{\tabcolsep}{2pt}
\renewcommand{\arraystretch}{1.13}
\begin{tabularx}{\linewidth}{@{}>{\hsize=0.56\hsize\linewidth=\hsize\raggedright\arraybackslash}X>{\hsize=0.86\hsize\linewidth=\hsize\raggedright\arraybackslash}X>{\hsize=0.42\hsize\linewidth=\hsize\raggedleft\arraybackslash}X>{\hsize=0.88\hsize\linewidth=\hsize\centering\arraybackslash}X>{\hsize=1.59\hsize\linewidth=\hsize\raggedright\arraybackslash}X>{\hsize=1.47\hsize\linewidth=\hsize\raggedright\arraybackslash}X>{\hsize=1.56\hsize\linewidth=\hsize\raggedright\arraybackslash}X>{\hsize=0.66\hsize\linewidth=\hsize\centering\arraybackslash}X@{}}
\toprule
\textbf{Model} & \textbf{Analysis role} & \textbf{Real n} & \textbf{Synthetic n (share)} & \textbf{Acoustics} & \textbf{Voice/delivery} & \textbf{Phrase variation} & \textbf{Main A--E stats?} \\
\midrule
A & Baseline & --- & --- & Unadapted; no synthetic training & Not applicable & Not applicable & Yes \\
\addlinespace[2pt]
B & Primary & 673 & 0 (0\%) & Real KDAB recordings only & Real speakers/\allowbreak{}rates & Observed authentic speech & Yes \\
\addlinespace[2pt]
C & Primary & 673 & 288 (30\%) & Full preset 7-profile processing & Full available TTS voice/\allowbreak{}delivery pool & Full structured phrase variability & Yes \\
\addlinespace[2pt]
D & Primary & 673 & 673 (50\%) & Same generation framework as C & Same framework as C & Same framework as C & Yes \\
\addlinespace[2pt]
E & Primary & 673 & 1,570 (70\%) & Same generation framework as C & Same framework as C & Same framework as C & Yes \\
\addlinespace[2pt]
G & Sensitivity & 673 & 288 (30\%) & Clean synthetic; broader degradation removed & As in C & Same 288 transcripts as C & No \\
\addlinespace[2pt]
H & Sensitivity & 673 & 288 (30\%) & Full radio-style processing & Reduced voice /\allowbreak{} accent-conditioned pool & Same 288 transcripts as C & No \\
\addlinespace[2pt]
I & Sensitivity & 673 & 288 (30\%) & Strong masking /\allowbreak{} interference removed; basic radio retained & As in C & Same 288 transcripts as C & No \\
\addlinespace[2pt]
J & Sensitivity & 673 & 288 (30\%) & Approx. matched to C* & Approx. matched to C* & New intent-matched subset; more canonical/\allowbreak{}complete & No \\
\bottomrule
\end{tabularx}
\par\smallskip
\begin{minipage}{\linewidth}
\footnotesize\raggedright
\textit{Note.} Models A--E were included in the principal Friedman and planned Wilcoxon comparisons; Models G--J were evaluated separately against Model C in exploratory sensitivity analyses. Models G--I reused Model C's 288 synthetic transcripts, whereas Model J used a separate intent-matched subset. A synthetic-only Model F was prepared but had no verified completed evaluation and was not included in the reported analyses.
\end{minipage}
\end{table}

All adapted Models B--E and G--J were initialized independently from OpenAI/Whisper-large-v3 and
trained using the same LoRA configuration: rank 16, alpha 32, dropout 0.05, no bias terms, and
target modules \texttt{q\_proj} and \texttt{v\_proj}. The rank-256, alpha-512 setting reported by
\citet{Arra2024} was not adopted directly because it was optimized through a dedicated
hyperparameter study on a substantially larger public ATC corpus of approximately 44,800 training
samples. Prior low-resource Whisper research also indicates that LoRA settings do not transfer
consistently across tasks, datasets, and model scales \citep{Wang2025}. Because the present study
examined training-data composition rather than LoRA optimization, a fixed lower-capacity
configuration was used across all adapted conditions; it was not claimed to be globally optimal for
KDAB.

Training used English transcription mode, 16-kHz Whisper input features, a learning rate of
\(10^{-4}\), 50 warm-up steps, per-device training and evaluation batch sizes of one, 16
gradient-accumulation steps, FP16 precision, and a maximum of 500 trainer steps. Validation and
checkpoint saving occurred every 50 steps, and generation length was limited to 128 tokens. The
checkpoint with the lowest WER on the common real validation set was selected for each run: step 100
for Model B, step 350 for Model C, step 200 for Model D, and step 400 for Model E. No synthetic
samples were used for validation or testing.

\subsection{Five-Seed Comparison of Real-Only and Augmented Training}

An independent repeatability assessment compared real-only Model B with the 30\% synthetic-augmented
Model C across five matched training seeds (11, 22, 33, 44, and 55). Model B used the 673 authentic
training transmissions; Model C used the same real transmissions plus the fixed subset of 288
synthetic examples. Both conditions used the same 145-transmission authentic validation set and the
same 145-transmission authentic test set. The runs used a separately versioned common protocol with
pinned Whisper Large-v3 initialization, identical LoRA settings, a maximum of 500 Trainer steps, and
checkpoint selection by authentic validation WER every 50 Trainer steps. Both models used the same
decoding procedure. The same normalization rules were applied to reference transcripts and decoded
hypotheses before scoring.

The five seed pairs were evaluated separately from the original single-run Models A--E comparisons.
Across the two series, differences in effect size cannot be attributed to seed choice alone.
Repeatability was summarized using seed-specific corpus WER/CER, across-seed means and sample
standard deviations, paired Model B-minus-Model C differences, and the number of seed pairs favoring
either condition. All seeds shared the fixed test transmissions and the same synthetic subset; the
experiment assessed training-seed variability rather than synthetic-sample selection or
between-airport generalization. Operational-field preservation was not rescored for the new runs.

\subsection{Ablation and Sensitivity Conditions}

Model C, the full 30\% structured synthetic condition, served as the reference for exploratory
component-sensitivity analysis (Table~\ref{tab:t2}). Models G, H, and I held the 673 authentic
training transmissions and the 288 synthetic transcripts fixed while altering the implemented
acoustic-treatment or voice/condition pool. Model G removed the broader synthetic radio/acoustic
degradation; Model H restricted voice and accent-condition diversity; Model I removed strong masking
and interference while retaining basic radio processing.

Phrase variability required a different design because it was encoded directly in the transcript.
Model J therefore used a newly selected 288-sample subset from the same 2,000-scenario pool. Its
intent-group distribution was matched to that of Model C, but the selection favored
canonical-complete structures and included fewer shortened or otherwise varied forms. Acoustic and
voice-accent conditions were matched as closely as the available pool allowed, although
communication-turn and speaker-role distributions could not be held constant. Model J was therefore
interpreted as an intent-matched sensitivity condition rather than a pure phrase-only ablation.

All ablation models used the same 673 real training segments, real validation set, fixed real test
set, Large-v3 initialization, LoRA configuration, and validation-based checkpoint-selection
procedure as Model C. Synthetic-ratio sensitivity was examined separately through Models C--E.

\subsection{Study Setting and Real KDAB Dataset}

The finalized real corpus contained 963 usable transmissions representing 58.73 minutes of speech.
As summarized in Table~\ref{tab:t3}, the corpus was divided into 673 training segments, 145
validation segments, and 145 fixed test segments. The split was performed at the
individual-transmission level using source-window stratification to preserve representation from the
recording periods across the three subsets.

\begin{table}[htbp]
\centering
\caption{Real KDAB Corpus Composition and Experimental Splits}
\label{tab:t3}
\small
\setlength{\tabcolsep}{4pt}
\renewcommand{\arraystretch}{1.15}
\begin{tabularx}{\linewidth}{@{}>{\hsize=1.12\hsize\linewidth=\hsize\raggedright\arraybackslash}X>{\hsize=0.71\hsize\linewidth=\hsize\raggedleft\arraybackslash}X>{\hsize=0.81\hsize\linewidth=\hsize\raggedleft\arraybackslash}X>{\hsize=1.15\hsize\linewidth=\hsize\raggedleft\arraybackslash}X>{\hsize=1.21\hsize\linewidth=\hsize\raggedleft\arraybackslash}X@{}}
\toprule
\textbf{Dataset subset} & \textbf{Segments} & \textbf{Duration (min)} & \textbf{Mean segment duration (s)} & \textbf{Median segment duration (s)} \\
\midrule
Training & 673 & 40.82 & 3.64 & 3.28 \\
Validation & 145 & 8.78 & 3.63 & 3.48 \\
Fixed test & 145 & 9.13 & 3.78 & 3.48 \\
\midrule
Total & 963 & 58.73 & 3.66 & 3.38 \\
\bottomrule
\end{tabularx}
\end{table}

Synthetic samples were added only to the training mixtures. The common 145-segment real validation
set was used for checkpoint selection, whereas the 145-segment real test set was excluded from model
updates and checkpoint selection. Because splitting was performed at the individual-transmission
level rather than by complete conversational exchange, temporally adjacent transmissions from the
same broader exchange could occur in different subsets.

\section{Evaluation metrics and statistical analysis}

All models were evaluated on the same fixed real KDAB test set. No synthetic samples were used for
validation or testing. Evaluation included transcript-level accuracy (WER, CER), segment-level
comparison, operational-information preservation, field-level preservation, qualitative examples,
and statistical tests.

\subsection{Transcript-Level Evaluation}

Transcript-level performance was evaluated using standard word error rate (WER) and character error
rate (CER) after applying the specified deterministic normalization to both references and
hypotheses. For each metric, the error rate is calculated as
\[
  \frac{\text{Substitutions}+\text{Deletions}+\text{Insertions}}
       {\text{Number of reference words (WER) or characters (CER)}}\,.
\]
Corpus-level WER/CER aggregate edit counts and reference-unit denominators across all 145 test
transmissions, whereas segment-level scores are used for paired comparisons.

For example, the reference is ``turn left heading one five zero,'' and the hypothesis is ``turn left
heading five zero.'' Deleting ``one'' produces \(\mathrm{WER} = \tfrac{1}{6} = 16.7\%\) and
\(\mathrm{CER} = \tfrac{4}{31} = 12.9\%\), when spaces are included in character-level counting.
However, the single-word deletion prevents the intended heading of \(150^{\circ}\) from being
accurately conveyed. In the second example, the reference is ``runway seven left,'' and the
hypothesis is ``runway seven.'' Deleting ``left'' produces \(\mathrm{WER} = \tfrac{1}{3} = 33.3\%\)
and \(\mathrm{CER} = \tfrac{5}{17} = 29.4\%\), including the deleted space. This error removes the
runway-side designator, demonstrating why transcript-level accuracy alone does not fully capture
operational-information preservation.

\subsection{Operational-Information Preservation}

A second evaluation layer assessed whether aviation-relevant information explicitly present in each
reference transcript was preserved in the ASR hypothesis. The final Models A--E comparison evaluated
intent type, full callsign, frequency, runway, taxiway, heading, handoff or facility, instruction
family, and clearance type. These fields differed from the callsign-core and strict command measures
used only in the preliminary architecture benchmark.

Reference targets and hypothesis evidence were compared using predefined aviation-specific
normalization, value, and phrase-family rules. Spoken and written equivalents, such as ``two five
right'' and ``25R,'' were mapped to a common representation, and each evaluable field was scored
independently within a segment. The analysis followed a strict no-inference rule: missing
information was not reconstructed from prior transmissions, expected phraseology, local airport
knowledge, or surveillance data. Only fields explicitly present in the reference entered the
corresponding denominator.

Operational performance was summarized using micro preservation across all evaluable field
instances, mean segment preservation, and field-specific preservation rates. Micro preservation
measured the proportion of reference fields retained across the corpus, whereas mean segment
preservation gave equal weight to each transmission. Field-level reporting was retained because
changes in WER or CER do not necessarily translate uniformly to callsign, runway, heading, intent,
or other operational fields.

\subsection{Qualitative Examples}

Quantitative evaluation was supplemented with illustrative transmission-level comparisons among the
unadapted, real-only, and 30\% synthetic-augmented conditions. Examples were selected to show the
recovery or loss of callsigns, runway and heading information, instruction or clearance families,
and local aviation terminology. These cases were used to explain observed error patterns and were
not treated as representative frequency estimates.

\subsection{Statistical Analysis}

Because the same 145 real test transmissions were evaluated under each model condition,
repeated-measures nonparametric tests were used. Friedman tests assessed overall differences in
segment-level WER, CER, and operational preservation. Planned two-sided Wilcoxon signed-rank tests
evaluated four contrasts for segment-level WER and operational preservation: Model A versus B, B
versus C, C versus D, and C versus E. Bonferroni correction for the four planned comparisons yielded
an adjusted significance threshold of \(\alpha=.0125\) within each outcome family. Field-specific
preservation rates were analyzed descriptively because the number of evaluable instances differed
substantially across operational categories.

Additional analyses used archived model predictions without retraining or new inference.
Minimum-edit-distance alignment was used to decompose the WER change between Models B and C into
substitution, deletion, and insertion differences. Model C was compared with each ablation condition
using paired Wilcoxon signed-rank tests with Holm adjustment, and matched rank-biserial effects
summarized the magnitude of the paired differences. The relationship between transmission difficulty
and the Model I--Model C WER difference was evaluated using Model B segment-level WER as a
difficulty proxy and Spearman's rho. Difficulty strata were used only to visualize this relationship
and were not treated as direct measures of acoustic noise or masking. Because the original
comparisons used one fitted model per condition, the paired tests quantify differences across the
fixed test transmissions conditional on those fitted models and do not estimate variability across
training seeds.

\section{Results}

\subsection{Transcript-Level Performance}

Table~\ref{tab:t4} reports transcript-level performance on the fixed 145-segment real KDAB test set.
The unadapted Large-v3 baseline, Model A, produced a WER of 0.6684 and a CER of 0.4833. Fine-tuning
on the 673 real KDAB training segments reduced these values to 0.2609 and 0.1755 for Model B,
corresponding to relative reductions of 61.0\% and 63.7\%, respectively.

Model C, whose training mixture contained 30\% synthetic audio, achieved the lowest error rates,
with a WER of 0.2188 and a CER of 0.1510. Relative to Model B, these values represented further
reductions of 16.1\% in WER and 14.0\% in CER. Models D and E both produced a WER of 0.2423, with
CERs of 0.1705 and 0.1702, respectively. Mean segment-level WER and CER showed the same overall
pattern, with Model C producing the lowest values.

\begin{table}[htbp]
\centering
\caption{Main ASR Performance on the Fixed Real KDAB Test Set}
\label{tab:t4}
\small
\setlength{\tabcolsep}{4pt}
\renewcommand{\arraystretch}{1.15}
\begin{tabularx}{\linewidth}{@{}>{\hsize=0.62\hsize\linewidth=\hsize\raggedright\arraybackslash}X>{\hsize=1.62\hsize\linewidth=\hsize\raggedright\arraybackslash}X>{\hsize=0.73\hsize\linewidth=\hsize\raggedleft\arraybackslash}X>{\hsize=0.73\hsize\linewidth=\hsize\raggedleft\arraybackslash}X>{\hsize=1.15\hsize\linewidth=\hsize\raggedleft\arraybackslash}X>{\hsize=1.15\hsize\linewidth=\hsize\raggedleft\arraybackslash}X@{}}
\toprule
\textbf{Model} & \textbf{Training condition} & \textbf{WER} & \textbf{CER} & \textbf{Mean segment WER} & \textbf{Mean segment CER} \\
\midrule
A & Unadapted Large-v3 & 0.6684 & 0.4833 & 0.7212 & 0.5333 \\
B & Real-only LoRA & 0.2609 & 0.1755 & 0.2897 & 0.1965 \\
C & Real + synthetic, 30\% & 0.2188 & 0.1510 & 0.2331 & 0.1619 \\
D & Real + synthetic, 50\% & 0.2423 & 0.1705 & 0.2677 & 0.1899 \\
E & Real + synthetic, 70\% & 0.2423 & 0.1702 & 0.2585 & 0.1823 \\
\bottomrule
\end{tabularx}
\end{table}

Segment-level WER differed across the five conditions, \(\chi^2(4,N=145)=344.525\), \(p<.001\).
Model B had lower segment-level WER than Model A (\(Z=-10.117\), \(p<.001\)), and, for the original
single-run fitted models, Model C had lower test-segment WER than Model B (\(Z=-4.000\),
\(p<.001\)). In the latter comparison, 66 segments improved under Model C, 23 worsened, and 56 were
tied. Model D had higher WER than Model C (\(Z=-2.793\), \(p=.005\)). Model E also had higher WER
than Model C, but the difference did not meet the Bonferroni-adjusted significance threshold
(\(Z=-2.334\), \(p=.020\)).

CER showed a comparable overall pattern, \(\chi^2(4,N=145)=308.881\), \(p<.001\). Model C had the
lowest CER mean rank (2.26), whereas Model A had the highest (4.85); Models B, D, and E occupied
intermediate positions.

A post hoc word-level alignment was used to characterize the WER reduction from Model B to Model C.
Across 1,828 reference words, total word errors decreased from 477 to 400. The reduction was
concentrated in substitutions, which decreased from 312 to 254, while insertions decreased from 70
to 50. Deletions remained essentially unchanged, increasing from 95 to 96 (Table~\ref{tab:t5}).

\begin{table}[htbp]
\centering
\caption{Word-Level Error Decomposition for Models B and C}
\label{tab:t5}
\small
\setlength{\tabcolsep}{5pt}
\renewcommand{\arraystretch}{1.16}
\begin{tabularx}{\linewidth}{@{}>{\hsize=1.5\hsize\linewidth=\hsize\raggedright\arraybackslash}X>{\hsize=0.88\hsize\linewidth=\hsize\raggedleft\arraybackslash}X>{\hsize=0.88\hsize\linewidth=\hsize\raggedleft\arraybackslash}X>{\hsize=0.74\hsize\linewidth=\hsize\raggedleft\arraybackslash}X@{}}
\toprule
\textbf{Error type} & \textbf{Model B} & \textbf{Model C} & \textbf{Change} \\
\midrule
Substitutions & 312 & 254 & \ensuremath{-}58 \\
Deletions & 95 & 96 & +1 \\
Insertions & 70 & 50 & \ensuremath{-}20 \\
\midrule
Total word errors & 477 & 400 & \ensuremath{-}77 \\
\bottomrule
\end{tabularx}
\end{table}

Overall, real-only adaptation produced the dominant reduction in transcript error, while the 30\%
synthetic mixture yielded a smaller but statistically significant additional WER reduction in the
original single-run comparison. The 50\% and 70\% mixtures did not produce lower error rates than
the 30\% condition.

\subsection{Operational-Information Preservation}

Operational-information preservation improved most after adaptation with real KDAB speech
(Table~\ref{tab:t6}). Micro preservation increased from 0.4915 for Model A to 0.6587 for Model B,
while mean segment preservation increased from 0.4674 to 0.6537. Model C achieved the highest
descriptive scores, with micro preservation of 0.6860 and mean segment preservation of 0.6811.
Models D and E each produced micro preservation of 0.6706, with mean segment preservation values of
0.6613 and 0.6719, respectively.

\begin{table}[htbp]
\centering
\caption{Operational-Information Preservation by Model}
\label{tab:t6}
\small
\setlength{\tabcolsep}{5pt}
\renewcommand{\arraystretch}{1.16}
\begin{tabularx}{\linewidth}{@{}>{\hsize=0.53\hsize\linewidth=\hsize\raggedright\arraybackslash}X>{\hsize=1.45\hsize\linewidth=\hsize\raggedright\arraybackslash}X>{\hsize=0.92\hsize\linewidth=\hsize\raggedleft\arraybackslash}X>{\hsize=1.1\hsize\linewidth=\hsize\raggedleft\arraybackslash}X@{}}
\toprule
\textbf{Model} & \textbf{Training condition} & \textbf{Micro preservation} & \textbf{Mean segment preservation} \\
\midrule
A & Unadapted Large-v3 & 0.4915 & 0.4674 \\
B & Real-only LoRA & 0.6587 & 0.6537 \\
C & Real + synthetic, 30\% & 0.6860 & 0.6811 \\
D & Real + synthetic, 50\% & 0.6706 & 0.6613 \\
E & Real + synthetic, 70\% & 0.6706 & 0.6719 \\
\bottomrule
\end{tabularx}
\end{table}

Segment-level preservation differed across the five conditions, \(\chi^2(4,N=145)=134.691\),
\(p<.001\). Mean ranks were 2.04, 3.17, 3.32, 3.23, and 3.23 for Models A through E, respectively.
Model B had higher preservation than Model A (\(Z=-6.372\), \(p<.001\)). The difference between
Models B and C was not statistically significant (\(Z=-1.337\), \(p=.181\); 25 segments improved
under Model C, 17 worsened, and 103 were tied). Model C also did not differ significantly from Model
D (\(Z=-1.190\), \(p=.234\)) or Model E (\(Z=-0.675\), \(p=.500\)).

Thus, real-only adaptation produced the principal improvement in operational-information
preservation, whereas the additional improvement observed for the 30\% synthetic condition was
descriptive and not statistically significant.

\subsection{Field-Level Preservation}

The aggregate preservation pattern did not extend uniformly across operational fields. From Model B
to Model C, instruction-family preservation increased from 0.615 to 0.688, and runway preservation
increased from 0.824 to 0.892. Intent type, taxiway, and handoff or facility preservation also
increased. In contrast, full-callsign preservation declined from 0.594 to 0.571, heading
preservation declined from 0.739 to 0.696, and clearance-type preservation remained unchanged at
0.612 (Table~\ref{tab:t7}).

\begin{table}[htbp]
\centering
\caption{Field-Level Preservation for the Real-Only and 30\% Synthetic Conditions}
\label{tab:t7}
\small
\setlength{\tabcolsep}{4pt}
\renewcommand{\arraystretch}{1.13}
\begin{tabularx}{\linewidth}{@{}>{\hsize=1.26\hsize\linewidth=\hsize\raggedright\arraybackslash}X>{\hsize=0.76\hsize\linewidth=\hsize\raggedleft\arraybackslash}X>{\hsize=1.12\hsize\linewidth=\hsize\raggedleft\arraybackslash}X>{\hsize=1.12\hsize\linewidth=\hsize\raggedleft\arraybackslash}X>{\hsize=0.74\hsize\linewidth=\hsize\raggedleft\arraybackslash}X@{}}
\toprule
\textbf{Operational field} & \textbf{Evaluable (n)} & \textbf{Model B} & \textbf{Model C} & \textbf{Change} \\
\midrule
Intent type & 127 & 0.638 (81/127) & 0.669 (85/127) & +0.031 \\
Full callsign & 133 & 0.594 (79/133) & 0.571 (76/133) & \ensuremath{-}0.023 \\
Frequency & 8 & 0.875 (7/8) & 1.000 (8/8) & +0.125 \\
Runway & 74 & 0.824 (61/74) & 0.892 (66/74) & +0.068 \\
Taxiway & 50 & 0.800 (40/50) & 0.840 (42/50) & +0.040 \\
Heading & 23 & 0.739 (17/23) & 0.696 (16/23) & \ensuremath{-}0.043 \\
Handoff/facility & 26 & 0.462 (12/26) & 0.500 (13/26) & +0.038 \\
Instruction family & 96 & 0.615 (59/96) & 0.688 (66/96) & +0.073 \\
Clearance type & 49 & 0.612 (30/49) & 0.612 (30/49) & 0.000 \\
\midrule
Total & 586 & 0.6587 (386/586) & 0.6860 (402/586) & +0.0273 \\
\bottomrule
\end{tabularx}
\end{table}

\subsection{Synthetic-Ratio and Ablation Results}

Among the three real-plus-synthetic mixture conditions, Model C produced the lowest WER and CER.
Increasing the synthetic share from 30\% to 50\% or 70\% did not reduce transcript error further on
the fixed real test set. Model D had significantly higher segment-level WER than Model C, whereas
the Model C--E difference did not meet the Bonferroni-adjusted significance threshold.

Table~\ref{tab:t8} compares Model C with four component-reduced conditions. Model I, which excluded
stronger masking and interference, showed the largest WER difference relative to Model C: Model I
returned a WER of 0.2899, an absolute increase of 0.0711 relative to Model C. The clean-synthetic
condition, Model G, returned a WER of 0.2566, while Model H, which used reduced voice and
accent-condition diversity, returned a WER of 0.2511. Models C, G, H, and I used the same 288
synthetic transcripts, and their CER results followed the same ordering.

\begin{table}[htbp]
\centering
\caption{Component-Level Sensitivity Relative to the Full 30\% Synthetic Condition}
\label{tab:t8}
\small
\setlength{\tabcolsep}{4pt}
\renewcommand{\arraystretch}{1.14}
\begin{tabularx}{\linewidth}{@{}>{\hsize=0.49\hsize\linewidth=\hsize\raggedright\arraybackslash}X>{\hsize=1.82\hsize\linewidth=\hsize\raggedright\arraybackslash}X>{\hsize=0.82\hsize\linewidth=\hsize\raggedleft\arraybackslash}X>{\hsize=0.82\hsize\linewidth=\hsize\raggedleft\arraybackslash}X>{\hsize=1.05\hsize\linewidth=\hsize\raggedleft\arraybackslash}X@{}}
\toprule
\textbf{Model} & \textbf{Synthetic condition} & \textbf{Test WER} & \textbf{Test CER} & \textbf{Absolute WER change vs. C} \\
\midrule
C & Full structured synthetic & 0.2188 & 0.1510 & --- \\
G & Clean synthetic & 0.2566 & 0.1855 & +0.0378 \\
H & Reduced voice/\allowbreak{}accent-condition diversity & 0.2511 & 0.1777 & +0.0323 \\
I & No strong masking/\allowbreak{}interference & 0.2899 & 0.1880 & +0.0711 \\
J & Low phrase variability & 0.2253 & 0.1549 & +0.0065 \\
\bottomrule
\end{tabularx}
\end{table}

Model J remained closest to Model C, with a WER of 0.2253 and a CER of 0.1549. Its separately
selected subset increased the proportion of canonical-complete phrases from 52.8\% to 80.2\% but
also increased the proportion of tower-side utterances from 59.0\% to 83.0\%. The Model C--J
comparison was therefore treated as an intent-matched sensitivity comparison rather than as a pure
phrase-only ablation.

Paired segment-level comparisons supported the descriptive ordering. After Holm adjustment, Model C
had lower WER than Model G (Holm-adjusted \(p=.000151\), matched rank-biserial
\(\text{effect}=0.528\)), Model H (Holm-adjusted \(p=.000338\), \(\text{effect}=0.525\)), and Model
I (Holm-adjusted \(p<.000001\), \(\text{effect}=0.644\)). The largest paired effect occurred in the
Model C--I comparison: Model C produced lower WER on 74 segments, higher WER on 19 segments, and
identical WER on 52 segments. Model J was not statistically distinguishable from Model C after
adjustment (Holm-adjusted \(p=.318\), \(\text{effect}=0.146\)).

\begin{figure}[htbp]
\centering
\includegraphics[width=0.86\linewidth]{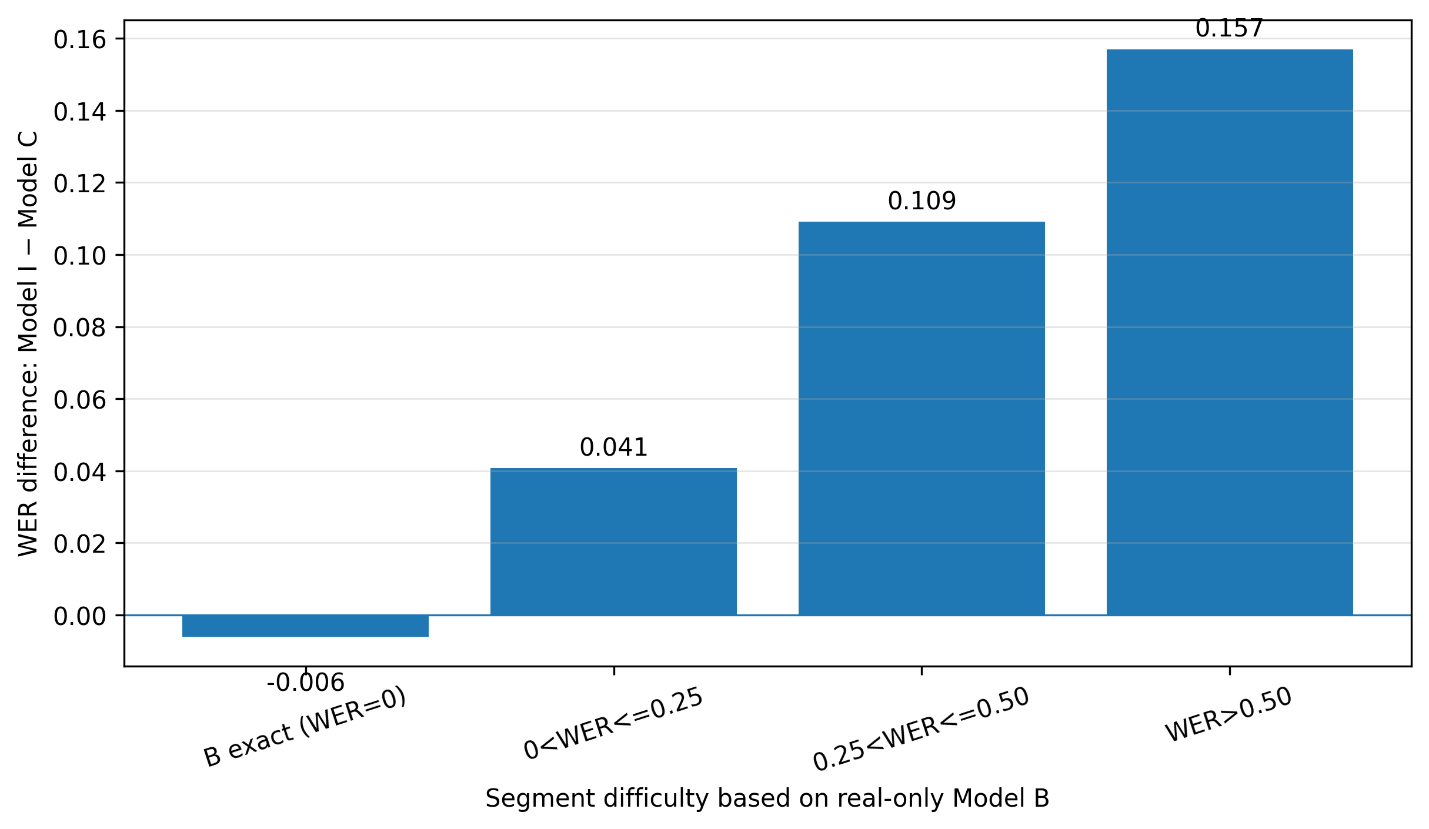}
\caption{Model C WER advantage over the no-strong-masking condition by baseline transmission difficulty.}
\label{fig:f3}
\end{figure}

Because Model I produced the largest paired difference, a follow-up analysis examined whether the
Model C--I difference varied across transmissions of different difficulty. Model B segment-level WER
was positively associated with the Model I--Model C WER difference (Spearman's \(\rho=.386\),
\(p<.001\)). The corpus-level WER difference was \(-0.006\) among the 16 segments transcribed
perfectly by Model B, 0.041 among the 65 segments with Model B WER greater than 0 and no more than
0.25, 0.109 among the 40 segments with WER greater than 0.25 and no more than 0.50, and 0.157 among
the 24 segments with Model B WER above 0.50. Positive values in Figure~\ref{fig:f3} indicate lower
WER for Model C. Model B WER was used as a model-based difficulty proxy rather than as a direct
measure of masking or acoustic quality.

The ratio comparison therefore identified the 30\% real-dominant mixture as the best-performing
tested share under the common training schedule, while the component comparisons associated the full
acoustic design---particularly the stronger masking and interference conditions---with lower WER.

\subsection{Qualitative Examples}

Table~\ref{tab:t9} presents five illustrative transmission-level comparisons across Models A, B, and
C. The first four examples show recovery of a callsign and handoff, a heading instruction, a landing
clearance, and a KDAB-local aviation entity. In these cases, Model C preserved complete operational
expressions that were missing or distorted in one or both comparison outputs. The final example
shows a remaining failure in which Model C produced a less accurate output than Model B,
demonstrating that synthetic augmentation did not improve every test transmission.

\begin{table}[htbp]
\centering
\caption{Illustrative ASR Outputs Across Models}
\label{tab:t9}
\footnotesize
\setlength{\tabcolsep}{3pt}
\renewcommand{\arraystretch}{1.14}
\begin{tabularx}{\linewidth}{@{}>{\hsize=0.72\hsize\linewidth=\hsize\raggedright\arraybackslash}X>{\hsize=1.1\hsize\linewidth=\hsize\raggedright\arraybackslash}X>{\hsize=0.98\hsize\linewidth=\hsize\raggedright\arraybackslash}X>{\hsize=1.1\hsize\linewidth=\hsize\raggedright\arraybackslash}X>{\hsize=1.1\hsize\linewidth=\hsize\raggedright\arraybackslash}X@{}}
\toprule
\textbf{Example} & \textbf{Reference} & \textbf{Model A} & \textbf{Model B} & \textbf{Model C} \\
\midrule
Callsign and handoff & [Callsign family A] four ninety three contact departure good day & Control 493 contact the fire trigger date & [Callsign family A] four nine three contact departure, good day & [Callsign family A]  four ninety three contact departure good day \\
\addlinespace[2pt]
Heading instruction & [Callsign family B] two zero seven turn left heading zero five zero & [Incorrect callsign] 207 turn left Phoenix 050 & [Callsign family B] two zero seven turn left ending zero five zero & [Callsign family B]  two zero seven turn left heading zero five zero \\
\addlinespace[2pt]
Landing clearance & cleared to land two five right & for the land 25 right & riddle and two five right & cleared to land two five right \\
\addlinespace[2pt]
Local aviation entity & sheltair & shelter & shelter & sheltair \\
\addlinespace[2pt]
Remaining failure & [Callsign family B] Seven Zero Seven Daytona Tower Roger & moving to 707, they do not downright & blue eight seven zero seven daytona tower roger & blue phoontop seven daytona ground \\
\bottomrule
\end{tabularx}
\par\smallskip
\begin{minipage}{\linewidth}
\footnotesize\raggedright
\textit{Note.} Callsign-family identifiers have been anonymized, and minor presentation normalization has been applied to selected transcripts.
\end{minipage}
\end{table}

\subsection{Five-Seed Comparison of Real-Only and Augmented Training}

Five paired B/C runs were completed under the common repeatability protocol. Across seeds, mean
corpus WER was \(0.2444\pm 0.0078\) for Model B and \(0.2357\pm 0.0091\) for Model C (mean \(\pm\)
sample SD); corresponding CER was \(0.1655\pm 0.0085\) and \(0.1570\pm 0.0080\), respectively
(Table~\ref{tab:t10}). The paired B-minus-C difference averaged 0.00875 in WER (0.875 percentage
points; \(\mathrm{SD}=0.01375\)) and 0.00844 in CER (0.844 percentage points;
\(\mathrm{SD}=0.01220\)). Model C achieved lower WER in three of five pairs and lower CER in four of
five; WER was higher for Model C in seeds 11 and 55.

\begin{table}[htbp]
\centering
\caption{Independent Five-Seed B/C Repeatability Assessment on the Fixed Real KDAB Test Set}
\label{tab:t10}
\small
\setlength{\tabcolsep}{3pt}
\renewcommand{\arraystretch}{1.16}
\begin{tabularx}{\linewidth}{@{}>{\hsize=0.82\hsize\linewidth=\hsize\raggedright\arraybackslash}X>{\hsize=1.045\hsize\linewidth=\hsize\raggedleft\arraybackslash}X>{\hsize=1.045\hsize\linewidth=\hsize\raggedleft\arraybackslash}X>{\hsize=1.045\hsize\linewidth=\hsize\raggedleft\arraybackslash}X>{\hsize=1.045\hsize\linewidth=\hsize\raggedleft\arraybackslash}X@{}}
\toprule
\textbf{Seed} & \textbf{Model B WER} & \textbf{Model C WER} & \textbf{Model B CER} & \textbf{Model C CER} \\
\midrule
11 & 0.2434 & 0.2451 & 0.1699 & 0.1663 \\
22 & 0.2527 & 0.2259 & 0.1705 & 0.1496 \\
33 & 0.2336 & 0.2303 & 0.1544 & 0.1513 \\
44 & 0.2511 & 0.2314 & 0.1741 & 0.1529 \\
55 & 0.2412 & 0.2456 & 0.1585 & 0.1651 \\
\midrule
Mean \ensuremath{\pm} SD & 0.2444 \ensuremath{\pm} 0.0078 & 0.2357 \ensuremath{\pm} 0.0091 & 0.1655 \ensuremath{\pm} 0.0085 & 0.1570 \ensuremath{\pm} 0.0080 \\
\bottomrule
\end{tabularx}
\par\smallskip
\begin{minipage}{\linewidth}
\footnotesize\raggedright
\textit{Note.} All five seeds used the same 145 authentic KDAB test transmissions and a fixed synthetic subset for Model C. Values represent corpus error rates; the final row reports the mean \ensuremath{\pm} sample standard deviation across five independently trained seeds.
\end{minipage}
\end{table}

The mean paired WER improvement under the repeatability protocol (0.875 percentage points) was
smaller than the 4.21-percentage-point difference in the original single-run B/C comparison. The
direction of the WER difference varied across training seeds. The repeated experiments followed a
common, separately versioned protocol and should not be interpreted as five direct reproductions of
the original historical model checkpoints or as independent test-data samples.

\section{Discussion}

The original single-run experiment produced three related findings. First, authentic KDAB speech
supplied the dominant adaptation signal. Second, a real-dominant mixture containing 30\% structured
synthetic speech achieved the lowest WER among the tested conditions. Third, the transcript-level
gain was selective: operational preservation improved only modestly, while full-callsign and heading
preservation remained uneven. Together, these findings show that the eight-layer generation
framework can serve as a structured supplement to authentic airport-specific training data.

\subsection{Airport-Specific Adaptation Under Limited Real Data}

The largest improvement in this study came from adaptation with authentic KDAB speech. Fine-tuning
Whisper Large-v3 on the airport-specific real corpus reduced WER from 0.6684 to 0.2609 and CER from
0.4833 to 0.1755. This result suggests that a relatively small but carefully curated corpus can
provide a strong adaptation signal when the target domain is narrow and locally repetitive.

Airport radio communication is constrained in ways that favor this type of specialization. Frequent
similar call-sign families, operational sequences, and channel characteristics recur across
transmissions. These regularities may not be represented with sufficient density in the broad
pretraining data of a general-purpose speech model. Exposure to authentic KDAB recordings likely
helped the model adjust not only to aviation terminology but also to the particular lexical and
acoustic patterns encountered at the airport.

The result should not be interpreted as evidence that less than one hour of speech is sufficient for
general ATC recognition. The model was adapted to a single airport and evaluated on recordings from
the same operational setting. Its improvement is more appropriately understood as narrow-domain
specialization. Larger and more diverse real corpora would still be needed to support recognition
across airports, facilities, and radio environments. This distinction also separates the present
study from work based on larger pooled ATC corpora and broader LoRA optimization. The purpose here
was not to identify a generally optimal parameter-efficient tuning configuration; it was to examine
whether limited airport-specific data could be supplemented with controlled synthetic variation.
Therefore, absolute performance should not be compared directly with studies using substantially
different data volumes and evaluation procedures.

\subsection{Substitutions and Insertions in the Original Model Comparison}

In the original single-run comparison, the 30\% synthetic condition yielded an additional reduction
in WER beyond that of real-only adaptation. The word-level decomposition clarifies the nature of
that improvement. As shown in Table~\ref{tab:t5}, Model C made fewer substitution errors and
insertion errors than Model B, while deletion errors were essentially unchanged. This indicates that
synthetic augmentation did not reduce all types of error equally.

The reduction in substitution is consistent with improved lexical selection. Structured synthetic
samples repeatedly presented operational phrases, local values, and plausible combinations of
aviation communication. This additional exposure may have made the model less likely to replace an
aviation term with a phonetically similar word from general language. The decrease in insertions
suggests that the model also became less likely to generate unsupported or extraneous words when
processing degraded radio speech. Deletion errors did not improve. This matters because a deletion
may reflect speech with weak acoustic evidence, be clipped, or be masked for the model to recover.
Synthetic exposure may broaden the model's expectation about what words are plausible, but it cannot
necessarily reconstruct information that is not sufficiently represented in the signal. The observed
pattern is therefore more consistent with improved selection and output control than with a general
ability to recover omitted speech.

This decomposition allows for a more specific interpretation of the transcript-level result. The
gain from synthetic augmentation was not merely a lower aggregate WER; it reflected fewer incorrect
word choices and fewer unnecessary additions. Also, the unchanged deletion count identifies a
remaining limit that may require stronger acoustic modeling and improved processing.

\subsection{Performance Across the Tested Training Mixtures}

Among the tested mixtures, the 30\% synthetic share produced the lowest WER and CER. Increasing the
synthetic proportion to 50\% or 70\% did not yield further improvement. This indicates that
synthetic speech was most useful as a controlled supplement rather than as an equal or dominant
source of training data under present conditions. This can be explained by the assumption of real
recordings acting as the primary target-domain anchor. The real recordings contained the actual
speakers, channel artifacts, and operational distributions that the model may encounter at
evaluation. The synthetic samples added combinations that were difficult to obtain from the small
real corpus alone. With the moderate share, these examples may have increased coverage without
displacing the authentic signal.

As the synthetic share increased, the model received a larger proportion of generated approximations
of the target environment. Even with carefully structured synthetic speech, it cannot reproduce
every feature of real tower communication. Differences may remain in domain variability and
operational characteristics. A synthetic-dominant mixture may expose the model more often to
regularities of the generation pipeline than to those of actual KDAB radio speech.

The training schedule may offer another plausible explanation. All mixture conditions were trained
to the same maximum number of steps, despite differing total dataset sizes. At a given step, the
larger mixtures had a smaller share of real examples. Validation also selected different checkpoints
for the real-only and 30\% models (100 and 350 steps, respectively). Therefore, the synthetic ratio
and the real-sample exposure were not fully disentangled. The result identifies 30\% as the
strongest ratio under the implemented schedule, not as a generally optimal proportion.

The separate five-paired-seed assessment provides a more qualified view of the additional gain from
synthetic augmentation. Model C had lower mean WER and CER than Model B, but the average WER
difference was only 0.875 percentage points and the WER ordering reversed in two seed pairs. This
contrasts with the 4.21-percentage-point historical single-run difference; the smaller effect cannot
be attributed to random seed alone because the repeatability runs used a separately standardized
protocol. As only Models B and C were repeated, the stability of the 50\% and 70\% mixture results
and the G--J sensitivity comparisons remains untested across seeds.

\subsection{What the Ablation Results Suggest}

The ablation results suggest that the benefit of synthetic augmentation depended on how the
synthetic speech was realized, not only on the transcript content. The clean-synthetic condition
performed worse than Model C despite using the same 288 transcripts. One plausible explanation is
that clean TTS encouraged the model to rely on overly regular acoustic patterns that were too far
removed from real tower audio. Radio filtering, cockpit background, microphone coloration,
compression, and related processing may have narrowed the mismatch by exposing the model to less
stable, less studio-like speech during training. The reduced voice and delivery diversity condition
may have produced a similar limitation. A narrower range of voices, speaking rates, roles,
hesitation patterns, and delivery styles could make it easier for the model to adapt to recurring
synthetic speaker characteristics rather than to features that remain useful across different
speakers and transmissions.

The strongest evidence concerned masking and interference. Model I retained the same transcripts as
Model C but removed the stronger masking condition, which produced the largest paired effect. Model
C showed almost no advantage in segments that Model B already transcribed perfectly. In contrast,
the gap widened as Model B's difficulty increased, reaching 0.157 for segments with Model B's WER
above 0.50. This pattern suggests that stronger interference was not uniformly beneficial. Instead,
it may have acted as a form of robustness training by requiring the model to recognize operational
speech when parts of the acoustic signal were weakened or obscured. Such exposure could reduce
reliance on consistently clear local cues and encourage greater use of partial acoustic and
linguistic context. Model B WER is a general proxy for difficulty, not a direct measure of noise,
masking, or signal quality. Difficult transmissions may also include rapid speech, unusual
pronunciation, rare local terms, atypical phrase structure, or multiple sources of difficulty
simultaneously. The analysis, therefore, provides evidence that the full masking condition was
particularly useful on difficult segments, but it does not establish which acoustic mechanism
produced the advantage.

The phrase-variability result may reflect the constrained structure of ATC communication. Even a
more canonical subset can repeatedly expose the model to the core combinations of callsigns,
runways, instructions, and readbacks needed for airport-specific recognition. This may explain why
Model J remained close to Model C. However, because Model J used a newly selected intent-matched
subset and did not hold communication turn and speaker-role distributions fully constant, the result
does not show that phrase variability was unimportant. More broadly, the ablation results should be
interpreted as a sensitivity pattern under the present dataset and implementation rather than as a
universal ranking of synthetic-data components.

\subsection{Operational Preservation Remains Uneven for Callsigns and Headings}

The operational information results qualify the transcript-level gains. Model C achieved the highest
descriptive micro and mean segment preservation scores, but its advantage over real-only fine-tuning
was modest and not statistically significant. Improvements were also uneven across fields. Full
callsign and heading preservation decreased, while clearance type remained unchanged; other fields
increased (Table~\ref{tab:t7}). This divergence is not contradictory. WER and CER respond to every
corrected word or character, regardless of its operational role. Operational preservation changes
only when a designated field is retained correctly. A model may reduce WER without correcting the
exact callsign or numeric value within the same transmission.

Runways and instruction families may have benefited from this because they occur within constrained
value sets and repeated phrase structures. KDAB runway identifiers recur throughout different
operations. Synthetic generation can reproduce these limited values and their surrounding phrases
repeatedly without requiring additional operational categories.

Full callsigns and headings present different recognition challenges. Callsigns contain variable
combinations of names, phonetic elements, and digits. If a single component is incorrect, the entire
field fails under exact scoring. Headings are short numeric sequences for which a single digit
changes the operational meaning. These fields remain sensitive to clipping, masking, pronunciation,
and confusion with nearby numbers. The mixed field-level result therefore shows why transcript
accuracy cannot serve as the sole evaluation criterion for aviation ASR.

The preliminary strict command accuracy result should be viewed in the same way. A command is passed
only when its intent and all required associated fields are correct together. A low strict score can
therefore result from cumulative failure across several components, even when the intent or runway
is recognized correctly. Such conjunctive measures are useful for exposing operational weakness, but
they should not be interpreted as equivalent to general intent-recognition accuracy or deployment
readiness.

\subsection{Implications for Communication Archives and ASR--ADS-B Reasoning}

Improved airport-specific ASR could support searchable tower-radio archives, post-operation review,
and communication-aware monitoring of runway and surface activity. These applications do not require
the system to replace controller judgment, but they would depend on reliable recovery of aircraft
identity and numeric instructions. In this context, the remaining instability in full-callsign and
heading recognition marks an important boundary: lower transcript error can make the communication
record more useful, but it does not, in itself, make the output suitable for safety-critical
automation.

The earlier KDAB multimodal framework \citep{Park2026} offers possible downstream use for these
improved transcripts. In that framework, ASR supplies callsign, runway, heading, and intent cues,
while ADS-B provides time-aligned position and kinematic evidence. More reliable speech recognition
could strengthen aircraft attribution, callsign recovery, and consistency checks between
communicated instructions and observed movement. The results establish an improved upstream ASR
component that could support future airport monitoring and decision-support applications.

\subsection{Eight-Layer Framework and National-to-Local Adaptation}

Beyond the observed ASR performance, the eight-layer framework provides a structured approach to
constructing synthetic ATC communication events and linking them to speech realization. The first
six layers organize operational intent, aircraft identity, operational values, airport-local
context, phraseology, and communication roles, while the final two layers specify speech delivery
and radio-acoustic processing. Maintaining these components under common sample-level identifiers
supports traceability between the intended communication, reference transcript, synthesis settings,
and processed audio. The present study evaluates the combined framework at KDAB rather than the
independent contribution of each layer.

A possible national-to-local adaptation strategy would maintain a shared foundation of ATC
communication intents, phraseological patterns, and operational-field definitions while allowing
individual airports to configure local runways, taxiways, recurring callsign patterns, and other
facility-specific information. Such a strategy could combine broader ATC speech resources with
verified local recordings and locally configured synthetic examples, then validate them on authentic
held-out communications at each target facility. This extension offers a conceptual direction for
scaling the framework beyond KDAB; the present study did not evaluate cross-airport transfer or
facility-specific acoustic calibration.

\subsection{Limitations and Future Work}

The findings are based on a single airport and a small real corpus totaling 58.73 minutes. Local
traffic, speakers, facilities, radio conditions, and callsign patterns may have contributed to the
observed performance. Multi-airport evaluation is needed to determine whether the same real-dominant
ratio and synthetic-component sensitivity also appear elsewhere.

The train, validation, and test sets were created at the individual-segment level. Adjacent
transmissions from a related exchange may therefore have been placed in different subsets. Although
the fixed test segments were excluded from fine-tuning and checkpoint selection, conversation-level
or recording-window-level separation would provide a stricter test of generalization.

Some transmissions in the fixed test set were also present in the earlier architecture-screening
benchmark.

They were not used to train the adaptation models, but the final test was not completely independent
of the prior decision to select Large-v3. The Models A--E comparison remains valid as a paired
comparison under a common architecture. In contrast, the absolute results should not be taken as
independent confirmation that Large-v3 is preferable to all alternative architectures.

The synthetic corpus was not evaluated through an independent, multi-rater perceptual study.
Therefore, we did not establish naturalness, realized accent validity, or perceived radio realism.
The actual voice-locale variation was centered on en-US and en-IN Standard voices, and metadata
categories that were not independently realized should not be interpreted as validated accent
conditions. Training-related limitations also constrain causal interpretation. LoRA hyperparameters
were held fixed rather than optimized for KDAB, and the common 500-step schedule confounded mixture
size with effective exposure to the real samples. The original mixture and component-sensitivity
results reflect one completed run per condition. The phrase-variability condition was an
intent-matched sensitivity comparison rather than a fully isolated ablation.

Although five paired training seeds were evaluated for Models B and C, the additional runs did not
assess synthetic-subset variability, independent recording sessions, or seed-level operational-field
preservation. The remaining principal mixture conditions and component-sensitivity models were not
repeated across seeds.

Finally, the masking--difficulty analysis was conducted post hoc, using Model B WER as a model-based
proxy for difficulty. It cannot determine which segments were difficult specifically because of
masking, noise, rate, pronunciation, lexical rarity, or operational complexity. Future work should
include predefined acoustic-quality measurements, completed repeatability analyses, equalized
real-data exposure, conversation-level splits, and tests across multiple airports. These additions
would allow more direct evaluation of the present sensitivity findings and stronger causal
separation.

\section{Conclusions}

This study examined whether structured synthetic ATC speech could extend a small airport-specific
corpus while preserving the value of authentic recordings. It introduced an eight-layer framework
linking operational communication-event construction to traceable speech synthesis and
radio-acoustic processing. In the original single-run experiment, fine-tuning Whisper Large-v3 on
673 real KDAB segments yielded the largest performance improvement, reducing WER from 0.6684 to
0.2609. Adding a 30\% synthetic share reduced WER further to 0.2188 and CER to 0.1510, whereas the
50\% and 70\% mixtures did not outperform the 30\% condition under the common training schedule.
These findings identify the 30\% real-dominant mixture as the best-performing tested condition in
the original comparison. In a separate five-paired-seed B/C assessment, the mean WER advantage was
smaller (0.875 percentage points) and was not observed in every seed pair.

The additional benefit was selective rather than uniform. Post hoc alignment showed that Model C
reduced substitutions and insertions, while deletions remained essentially unchanged. The same model
achieved the highest descriptive operational-preservation scores, but its advantage over real-only
adaptation was not statistically significant, and full-callsign and heading preservation did not
improve. Component-level analyses further associated the full acoustic design, particularly the
masking and interference conditions, with lower WER on transmissions that the real-only model found
difficult. These comparisons identify sensitivities of the present design; they do not establish a
universal ranking of synthetic components or a generally optimal synthetic ratio.

Within these boundaries, the findings support a real-first approach to low-resource airport ASR:
authentic recordings define the local domain, while structured synthetic audio broadens controlled
voice, delivery, phraseological, and channel variation around that anchor. This strategy may be
relevant when access, sharing, or annotation constraints limit proportional expansion of real
airport corpora, although those burdens were not measured directly. Broader application requires
conversation- or time-blocked evaluation, repeated training seeds, equalized real-sample exposure,
independent perceptual assessment of synthetic speech, and testing at additional airports. Improved
recognition may eventually strengthen communication archives, airport monitoring, and multimodal
ASR--ADS-B reasoning, but the present study establishes an upstream adaptation benefit rather than
operational readiness or autonomous ATC capability.

\section*{Acknowledgments}

The authors used OpenAI ChatGPT, including GPT-5.5 and GPT-5.6 Thinking, during June--July 2026 to
assist with language editing, development and review of analysis code, and the conceptual design and
visual layout of Figure~\ref{fig:f1}. ChatGPT was not used to generate or alter the study data or to
independently execute the reported experiments. All AI-assisted text, code, calculations, citations,
interpretations, and figure elements were reviewed, verified, and revised by the authors, who take
full responsibility for the final manuscript. ChatGPT also assisted with editing and formatting the
preprint in November 2026.

\section*{Author contributions}

The authors confirm contribution to the paper as follows: study conception and design: J. Park, C.
Yang; data collection: J. Park; analysis and interpretation of results: J. Park, C. Yang; draft
manuscript preparation: J. Park, C. Yang; funding acquisition: C. Yang.

\section*{Declaration of conflicting interests}

The authors declared no potential conflicts of interest with respect to the research, authorship,
and publication of this article.

\section*{Funding}

This work was supported by the Embry-Riddle Aeronautical University FIRST Grant (GC 87629).

\FloatBarrier


\end{document}